# Development of Autonomous Artificial Intelligence Systems for Corporate Management

Anna Romanova

*Aspirant, MIPT*
*Russian Federation, Saint Petersburg*

Abstract

The article discusses development of autonomous artificial intelligence systems for corporate management. The function of a corporate director is still one of the few that are legislated for execution by a "natural" rather than an "artificial" person. The main prerequisites for development of systems for full automation of management decisions made at the level of a board of directors are formed in the field of corporate law, machine learning, and compliance with the rules of non-discrimination, transparency, and accountability of decisions made and algorithms applied. The basic methodological approaches in terms of corporate law for the "autonomous director" have already been developed and do not get rejection among representatives of the legal sciences. However, there is an undeniable need for further extensive research in order to amend corporate law to effectively introduce "autonomous directors". In practice, there are two main options of management decisions automation at the level of top management and a board of directors: digital command centers or automation of separate functions. Artificial intelligence systems will be subject to the same strict requirements for non-discrimination, transparency, and accountability as "natural" directors. At a certain stage, autonomous systems can be an effective tool for countries, regions, and companies with a shortage of human capital, equalizing or providing additional chances for such countries and companies to compete on the global market.

**Keywords:** artificial intelligence, corporate governance, autonomous, machine learning, board of directors

# Развитие автономных систем искусственного интеллекта для управления корпорациями

Романова Анна Сергеевна

*Аспирант, МФТИ*
*Российская Федерация, Санкт-Петербург*

**Аннотация**

В статье рассматривается развитие автономных систем искусственного интеллекта для управления корпорациями. Функция корпоративного директора пока еще является одной из немногих, законодательно закрепленных для исполнения именно "натуральным", а не "искусственным" лицом. Основные предпосылки развития систем полной автоматизации управленческих решений, принимаемых на уровне совета директоров, формируются в области корпоративного права, машинного обучения, и соблюдения правил недискриминации, прозрачности, и подотчетности принимаемых решений и применяемых алгоритмов. На данный момент, базовые методологические подходы в части корпоративного права для "автономного директора" уже проработаны и не встречают отторжения среди представителей правовых наук. Тем не менее существует бесспорная необходимость в дальнейших обширных исследованиях с целью внесения изменений в корпоративное право для эффективного внедрения "автономных директоров". На практике сложились два основных направления автоматизации управленческих решений на уровне топ-менеджмента и совета директоров: цифровые командные центры или автоматизация отдельных функций. Какие бы функции не выполняла система искусственного интеллекта, к ней будут применяться столь же строгие требования недискриминации, прозрачности, и подотчетности, что и к обычным ("натуральным") директорам. На определенном этапе автономные системы могут оказаться эффективным инструментом для стран, регионов, и компаний с дефицитом человеческого капитала, уравнивая или предоставляя дополнительные шансы таким странам и компаниям в конкуренции на мировом рынке.

**Ключевые слова:** искусственный интеллект, корпоративное управление, автономность, машинное обучение, совет директоров

**Введение**

В отчете Всемирного экономического форума (WEF) за 2015 год "Переломные моменты технологий и влияние на общество" представлен прогноз, что уже к 2026 году первая система с искусственным интеллектом (далее ИИ) займет место в корпоративном совете директоров [39]. Первое официальное объявление о работе системы искусственного интеллекта в совете директоров было опубликовано еще в 2014 году [6]. Гонконгская венчурная компания Deep Knowledge Ventures заявила о назначении системы VITAL (Validating Investment Tool for Advancing Life Sciences) членом своего совета директоров [6]. Скандинавская компания Tieto в 2016 году объявила о назначении системы ИИ Alicia T топ-менеджером новой бизнес-единицы [7]. Alicia T стала первой системой, назначенной управлять публичной компанией, котирующейся на Nasdaq Nordic.

Публичные назначения систем ИИ в советы директоров (далее СД) в настоящее время носят в основном рекламный характер, но уже существует обширная практика использования систем на базе алгоритмов машинного обучения для подготовки принимаемых топ-менеджментом решений. Как отмечает профессор Университета Южной Калифорнии Алекс П. Миллер (Alex P. Miller), в настоящее время происходит тихая революция, которая характеризуется неуклонным ростом автоматизации традиционно принимаемых человеком решений в организациях [23].

Официальное принятие законодательства, позволяющего системе ИИ официально исполнять функции корпоративного директора, будет является важной вехой в процессе цифровизации общества. Функция корпоративного директора является одной из немногих, законодательно закрепленных для исполнения именно "натуральным", а не "искусственным" лицом. Основные предпосылки развития систем полной автоматизации управленческих решений, принимаемых на уровне совета директоров, формируются в области корпоративного права, машинного обучения, и соблюдения правил недискриминации, прозрачности, и подотчетности принимаемых решений и применяемых алгоритмов.

**Современные основы корпоративного управления (corporate governance)**

Одним из общепризнанных мировых стандартов корпоративного управления являются "Принципы корпоративного управления G20/ОЭСР" [26]. Принципы "определяют ключевые элементы структуры корпоративного управления и предлагают практическое руководство для их применения на национальном уровне" [26]. Шестой раздел Принципов раскрывает основные обязанности совета директоров. Принципы не устанавливают требование исполнения обязанностей директора именно физическим лицом. В современной корпоративной практике существует концепция корпоративного директора, когда "юридические лица выступают в качестве директоров компании" [4].

Принципы ОЭСР устанавливают: "Структура корпоративного управления должна обеспечивать стратегическое руководство компанией, эффективный контроль за менеджментом со стороны совета директоров, а также подотчетность совета директоров перед компанией и акционерами" [26].

"Основными функциями, которые должен выполнять СД являются:

- Пересматривать и направлять корпоративную стратегию, основные планы действий, политику и процедуры управления рисками, годовые бюджеты и бизнес- планы, определять целевые результаты деятельности; осуществлять контроль за выполнением планов и деятельностью корпорации, а также контролировать крупные капитальные расходы, покупки и продажи" [26].

- "Контролировать эффективность практики управления компанией и вносить в нее изменения по мере необходимости" [26].

- "Подбирать ключевых руководящих лиц, назначать им оплату, осуществлять контроль за их деятельностью и, в случае необходимости, заменять их и следить за планированием кадрового обновления" [26].

- "Устанавливать вознаграждение, получаемое ключевыми руководящими лицами и членами совета директоров, в соответствии с долгосрочными интересами компании и её акционеров" [26].

- "Обеспечивать соблюдение формальностей и прозрачности в процессе выдвижения и избрания членов совета директоров" [26].

- "Контролировать и урегулировать потенциальные конфликты интересов менеджмента, членов совета директоров и акционеров, включая ненадлежащее использование активов корпорации и злоупотребления при совершении сделок со связанными сторонами" [26].

- "Обеспечивать целостность систем бухгалтерской и финансовой отчетности, включая независимый аудит, а также наличие соответствующих систем контроля, в частности, систем управления рисками, финансового и операционного контроля, а также соблюдения законодательства и соответствующих стандартов" [26].

- "Осуществлять надзор за процессом раскрытия информации и коммуникаций" [26].

Требования, кто и каким образом должен исполнять обязанности директора, Принципы считают прерогативой норм локального законодательства каждой страны. Таким образом, если нормы локального законодательства позволяют использовать систему ИИ в качестве директора, а имеющиеся технологии поддерживают фактическое выполнение необходимых функций директора, то корпорация легитимно может назначить автономного директора в СД.

Развитие правовых основ применения систем искусственного интеллекта для управления корпорациями

Профессор Калифорнийского университета Меир Дан-Коэн (Meir Dan Cohen) в 1970 году предложил концепцию полностью автоматизированной корпорации - "personless corporation" [10]. По мнению Дан-Коэна, замена менеджмента корпорации на компьютеры будет иметь минимальный эффект на операционную деятельность и юридический статус корпорации. Дан-Коэн однако указывает, что необходимым условием является именно успешный переход на автоматизированное принятие решений. В случае успеха, юридический статус корпорации не изменится. По мнению Дан-Коэна полностью автоматизированная корпорация достаточно легко пройдет своеобразный тест Тьюринга для корпораций: прибыль будет реинвестирована, дружественные политики и культурные события будут своевременно спонсироваться [10]. Дан-Коэн указывает, что автономная корпорация должна будет иметь права. Таким образом, необходимо решение проблемы – может ли полностью автономная корпорация иметь права. Дан-Коэн предсказывает, что ответ будет положительным [10].

Теоретические идеи Дан-Коэна получили дальнейшее развитие в XXI веке в связи с успешным развитием технологий. Реализуемый практически подход продемонстрировал профессор Университета штата Флорида Шон Баверн (Shawn Bayern) в цикле статей посвященном автономным компаниям [5]. Баверн показал, что любой может присвоить статус юридического лица автономному компьютерному алгоритму путем передачи его под контроль компании с ограниченной ответственностью. Профессор Калифорнийского университета в Лос-Анджелесе Линн ЛоПаки (Lynn LoPucki) в продолжение работ Шона Баверна вводит понятие алгоритмической компании ("algoritmic entity"), когда алгоритм будет исполнять все права и обязанности юридического лица [22]. ЛоПаки также показал, что компания, созданная по законам штата Делавэр необязательно должна управляться менеджером- человеком, а может управляться искусственным лицом [22]. ЛоПаки проанализировал текущую юридическую готовность к возникновению алгоритмических компаний в нескольких юрисдикциях. Анализ выявил, что даже без модификации текущего корпоративного законодательства уже возможны юридические конструкции для работы алгоритмических компаний [22].

Вместе с анализом практических возможностей развивалась и применяемая терминология, а также инструменты для создания автономного СД. Если Дан-Коэн использовал термин "безличная корпорация" (personless corporation), то в 2019 году профессор права и финансов Оксфордского университета Джон Армор (John Armour) и профессор коммерческого права Оксфордского университета Хорст Эйденмюллер (Horst Eidenmuller) вводят термин автономная/беспилотная (self-driving) компания – по аналогии с беспилотным автомобилем (self-driving car) [22]. Также в 2019 году профессор Калифорнийского университета в Лос-Анджелесе Мартин Петрин (Martin Petrin) ввел понятие компании без лидеров ("leaderless entity") и развил понятие алгоритмической компании [27]. Уже существуют алгоритмы, которые могут существовать автономно. Самый простой пример – это компьютерные вирусы. Петрин предполагает, что продвинутые формы таких алгоритмов могут самостоятельно вести бизнес. Петрин рассматривает вопрос, а будет ли в таком случае существовать необходимость в корпоративном менеджменте в принципе? Петрин предлагает концепцию новых бизнес-структур, которые будут функционировать без лидерства в традиционном смысле. Петрин приводит пример компании ConsenSys, компании по разработке программного обеспечения, специализирующейся на приложениях для блокчейн платформы Ethereum [27].

Профессор Петрин также вводит понятие смешанного совета директоров ("fused board"): "где различные роли и вклады, ранее предоставляемые коллективом директоров-людей, включены в единое программное обеспечение или алгоритм, производительность которого превзойдет сегодняшнее управление, предоставляемое людьми" [27]. Петрин считает, что в дальнейшем: "...ИИ также заменит управленцев и менеджеров ниже уровня СД. ...эти разработки в конечном итоге сделают разделение СД и менеджмента устаревшим и приведут к объединенному управлению ("fused management") корпорациями, при этом компании будут полностью управляются одним подразделением ИИ" [27].

Профессор Марбургского университета Флориан Мослейн (Florian Moslein) связывает переход к автономному ИИ управлению прежде всего с возникновением децентрализованных автономных организаций ("decentralized autonomous organizations", DAOs): "Эти организации управляются в соответствии с правилами, закодированными в виде компьютерных программ (так называемые смарт-контракты); эти правила, а также записи их транзакций, хранящиеся в блокчейне, означают, что они могут работать полностью без участия человека" [25]. Мослейн считает, что для эффективной работы смешанного СД или смешанного менеджмента потребуется ввести правила распределения прав принятия решений между человеком и машиной [25]. Профессор Мослейн использует понятие робо-директора ("robo-director") [Moslein]. Жиан Моско (Gian Mosco), профессор университета имени Гвидо Карли, по аналогии с термином Илона Маска (Elon Musk) "роботакси" ("robotaxi") предлагает термин "робосовет" ("roboboard"), который относится к СД, состоящему из искусственных директоров [24].

В 2019 году профессор Оксфордского университета Люка Энрикес (Luca Enriques) и профессор Люксембургского университета Дирк Цетше (Dirk Zetzsche) вводят общий термин CorpTech [11]. В формулировке Энрикеса и Цетше: "CorpTech включает в себя все решения относящиеся к корпоративному управлению в широком смысле, включая инструменты для установления компенсаций, выявления кандидатов на высокие должности в организации, поддерживания отношений с инвесторами, процедуры по корпоративному голосованию и внутренней работе совета директоров, управлению рисками, и улучшению эффективности функции СД" [11].

Таким образом, на данный момент, базовые методологические подходы в части корпоративного права для "автономного директора" уже проработаны и не встречают отторжения среди представителей правовых наук. Тем не менее существует бесспорная необходимость в дальнейших обширных исследованиях с целью внесения изменений в корпоративное право для эффективного внедрения "автономных директоров".

**Использование технологий искусственного интеллекта для управления корпорациями**

Еще в 1965 году один из пионеров искусственного интеллекта профессор Университета Карнеги - Меллона в Питтсбурге Герберт Саймон (Herbert Simon) сформулировал концепцию автономной фабрики и концепцию автономной корпорации [35]. Симон предсказывал автоматизацию почти всей работы клерков, что и произошло на самом деле, а также предсказывал возникновение автоматизированных топ- менеджеров ("automated executive"). В его представлении "сложные системы обработки информации" смогут заменить людей в задачах, где надо "думать и принимать решения". Топ-менеджеры, в понимании Саймона, будут заняты только в тех задачах, где необходимо эмоциональное взаимодействие с работниками. Саймон ожидал, что все основные перемены уже произойдут к 1985 году [35].

В отчете 2016 года международная консалтинговая компания Accenture рассматривает переход систем ИИ от пассивной роли в управлении компанией к активной за три основных этапа:

- ИИ как ассистент - создает скоринговые карты, ведет отчеты, делает заметки, управляет расписание, мониторит среду, отслеживает исполнение принятых решений;

- ИИ как советник - отвечает на вопросы, разрабатывает сценарии, генерит опции, ведет встречи, анализирует поведение команды, рекомендует роли в команде;

- ИИ как действующее лицо - оценивает опции, принимает решения, ведет бюджетирование и планирование, оценивает динамику команды, меняет статус-кво [36];

В настоящее время на практике сложились два основных направления автоматизации управленческих решений на уровне топ-менеджмента и СД: командные центры или автоматизация отдельных функций. Многие компании переходят к автоматизации принятия решений на уровне топ-менеджмента не формализуя такую структуру, как центр принятия ключевых решений компании, но создавая на практике цифровые командные центры. Также развиваются отдельные системы ИИ, которые успешно исполняют некоторые ключевые функции СД.

"Командные центры существуют уже очень давно; крупные организации и правительства использовали их с самого начала цивилизации. Однако технологии, которые позволяют командным центрам выполнять свои задачи, сильно изменились, и для обеспечения успеха требуются новые навыки и подходы" [31].

Одним из современных примеров является цифровой командный центр (ЦКЦ) нефтяной компании ADNOC, введенный в эксплуатацию несколько лет назад. Цифровой командный центр Panorama объединяет информацию в режиме реального времени по более чем дюжине дочерних и совместных предприятий. В системе также используется интеллектуальная аналитическая модель и искусственный интеллект для создания оперативных выводов и рекомендаций [17]. ЦКЦ Panorama на практике демонстрирует как реализуются теоретические дискуссии об изменении парадигмы топ-менеджмента в эпоху ИИ: снижение агентских издержек, прозрачность и подотчетность информации, отслеживание данных в реальном времени и т. д. В ЦКЦ Panorama для мониторинга используется технология цифровых двойников, для записи и сохранения информации используется технология блокчейн. В сущности, ЦКЦ Panorama представляет собой цифровую фабрику, которая обрабатывает информацию и предлагает рекомендации в рамках компании с оборотом более 60 миллиардов долларов США [17].

Одной из базовых технологий, которая позволяет получать информацию по всей компании в реальном времени, является технология цифровых двойников ("digital twins"). Технология возникла в начале текущего столетия и была направлена на то, чтобы создать цифровую модель физической системы перед ее построением. Затем технология соединилась с сенсорами интернета вещей и в результате стало возможным "...добавить больше данных, чтобы обогатить цифрового двойника, собирая данные об окружающей среде, такие как местоположение и конфигурация, а также более общую информацию, такую как записи об услугах, финансовые модели и т. д." [15]. В результате: "...организации в итоге получают цифровую конструкцию, которая знает все о проектировании, строительстве или производстве, прошлых и текущих операциях и обслуживании физической системы. Если добавить интеллект в виде аналитики, моделей и других алгоритмических методов, таких как машинное обучение, организации смогут начать получать прогнозы и ранние предупреждения быстрее, чем когда-либо прежде" [15].

Еще одной технологией, которая позволила улучшить прозрачность и достоверность управленческих решений, стала технология блокчейн. Современная технология блокчейн стала известной в 2008 году, когда она была использована для создания криптовалюты биткойн [40]. Блокчейн служит неизменным регистром, который позволяет проводить транзакции децентрализованным образом. Таким образом, принимая управленческое решение, которое сформировано на информации, записанной в блокчейн, можно ожидать, что информация достоверна и прозрачна.

Третьим, важным компонентом, который позволил создать "операционный цифровой мозг" в промышленном масштабе, стали технологии машинного обучения. Самые простые формулы, применяемые в машинном обучении, возникли много лет назад, но значительнее всего технологии машинного обучения продвинулись "за последние два десятилетия" [19]. "Недавний прогресс в машинном обучении был обусловлен как разработкой новых алгоритмов обучения и теории, а также продолжающимся взрывом в доступности онлайн-данных и недорогих вычислений" [19].

В настоящее время сформированы и правовые, и технологические предпосылки для создания полностью автоматизированной системы управления корпорацией в виде цифровой фабрики. Основными проблемами, сдерживающими массовое внедрение являются слишком высокая стоимость, недостатки законодательства, психологическая неподготовленность основных заинтересованных лиц и необходимость избегать социальной напряженности, связанной с возможной потерей рабочих мест.

Многие исследователи и компании идут также по пути создания систем ИИ, которые выполняют отдельные функции СД и далее могут быть объединены в "fused boards".

Система ИИ, которая исполняет функции ассистента на заседаниях топ-менеджмента была представлена в 2018 году на встрече WEF в Давосе [27]. Это система Einstein, которую CEO компании SalesForce Марк Бениофф (Marc Benioff) использует на еженедельных встречах сотрудников. Бениофф привел пример, когда при обсуждении планов Einstein смог более объективно оценить обстановку, чем высококвалифицированный управленец: "«У меня за столом 30 или 40 топ-менеджеров, — сказал он (Бениофф). «И мы выясняем, как у нас дела, когда смотрим на весь этот анализ». Бениофф сказал, что после того, как он поработал в своей комнате с руководителями, он затем опрашивает своего помощника по искусственному интеллекту. «Я спрашиваю Эйнштейна: «Я слышал, что все говорили, но что вы на самом деле думаете?» Руководитель Salesforce сказал, что использует эту технологию уже год, а недавно Эйнштейн подверг сомнению отчеты одного европейского сотрудника. «И Эйнштейн сказал: «Ну, я не думаю, что этот руководитель добьется успеха, мне очень жаль»" [28].

Чтобы создать платформу предиктивной аналитики Einstein Discovery, SalesForce купила несколько стартапов и создала масштабную команду специалистов по машинному обучению. В итоге "...большинство реальных усилий, выполненных Einstein Discovery сфокусированы на повышении прозрачности бизнес-данных, создании интуитивно понятных моделей, и практическом применение прогнозов и действенных идей лицами, принимающими бизнес-решения" [33].

Предиктивная аналитика все больше используется в автоматизации решений, которые раньше принимались CEO или СД компании, например, дивидендная политика. Дивидендная политика означает решение о проценте выплаты дивидендов от прибыли, выплачиваемых акционерам. Она касается суммы и сроков денежных выплат акционерам компании. Решение о дивидендной политике является очень важным для фирмы, поскольку оно может повлиять на структуру ее капитала и цену акций. В 2010 году исследователи из Университета Кемен в Южной Корее опубликовали результаты определения оптимальной дивидендной политики с помощью алгоритмов машинного обучения. Для тренировки алгоритмов они использовали публично доступные данные с биржи и продемонстрировали эффективную работу алгоритмов для поддержки решений о дивидендах [3].

Еще одной функцией, которая в настоящее время начала автоматизироваться, является функция подбора и назначения руководителей. Алгоритмы машинного обучения уже активно используются для подбора персонала низшего и среднего звена, но их использование для подбора топ-менеджеров еще не стало мейнстримом. В СД директора обычно работают совместно, поэтому показатели, которые характеризуют работу каждого отдельного директора, напрямую недоступны.

В 2021 году группа ученых из Университета штата Огайо, Колорадского университета, и Вашингтонского университета опубликовала результаты исследования о том, как машинное обучение может быть использовано для выбора членов совета директоров и чем выбранные директора могут быть эффективнее тех, которые были выбраны действующим менеджментом компании [12]. Поскольку при назначении директора СД должен сделать прогноз эффективности директора, то исследователи сформулировали задачу для алгоритмов машинного обучения как задачу предсказания. В данном случае для эксперимента ученые сформировали специальную базу данных публичных американских фирм и независимых директоров, назначенных в период с 2000 по 2014 год, а также сформировали цифровые показатели (переменные), которыми можно характеризовать директора, СД, и компанию.

Очень важным направлением работы СД является взаимодействие с акционерами. В 2018 году исследователи из Манчестерской школы бизнеса опубликовали результаты эксперимента по использованию систем ИИ в качестве ассистента в процессе принятия решений человеком с целью активизации участия акционеров в работе компании. Решения общего собрания акционеров могут включать:

- Утверждение годового отчета.

- Утверждение предлагаемой структуры вознаграждения менеджмента.

- Утверждение назначения и вознаграждения аудиторов.

- Утверждение назначения избранного состава правления, включая председателя.

- Вопросы, касающиеся выпуска акций.

- Различные вопросы, такие как право на проведение общего собрания акционеров с уведомлением, политические пожертвования [Carpenter].

Используя исторические и публичные данные компании, ученые представили прототип ассистента на базе алгоритмов машинного обучения и обработки текстов на естественном языке, для поддержки принятия решений по голосованию на годовом общем собрании акционеров [8].

В настоящее время также разрабатываются специальные модули для поддержки работы СД на базе действующих ERP систем, в частности, SAP Digital Boardroom. Действующая версия уже поддерживает следующие функции:

- Интеллектуальные встречи (поддерживает анализ в режиме реального времени, предоставляет интерактивные визуализации данных и адаптивных макетов).

- Мгновенная аналитика на основе данных.

- Планирование и моделирование (моделирование влияния решений с помощью специального анализа и анализа «что, если»).

- Передовой бизнес-контент (подготовленные модели данных, визуализаций, и лучших практик) [30].

Таким образом, собирая вместе отдельные автоматизированные функции СД, в итоге можно создать рассмотренный ранее смешанный совет директоров ("fused board"), "где различные роли и вклады, ранее предоставляемые коллективом директоров-людей, включены в единое программное обеспечение" [27].

**Внедрение правил недискриминации, прозрачности и подотчетности для систем искусственного интеллекта**

Какие бы функции СД не выполняла система ИИ, к ней будут применяться столь же строгие требования недискриминации, прозрачности, и подотчетности, что и к обычным ("натуральным") директорам. Современные исследователи предлагают три основные походы к управлению рисками, возникающими при использовании моделей искусственного интеллекта: прозрачность, объяснимость, и подотчетность [18]. Исследователи выделяют несколько этапов разработки системы ИИ, где могут возникать ошибки предвзятости "входные данные, обучение, и программирование" [18]. В настоящее время развивается несколько подходов, имеющих своей целью выработать методы исключения ошибок систем ИИ. Основное отражение эти подходы находят либо в государственном регулировании процесса разработки и систем использования ИИ, либо в использовании определённых технологий.

В частности, "Европейский союз пытается решить проблему «черного ящика» и вытекающую из этого проблему прозрачности, предоставляя гражданам ЕС так называемое «право на объяснение»" [14]. Предполагается, что введение "права на объяснение" "...заставит разработчиков по-другому структурировать алгоритмы и системы, чтобы обеспечить более высокую степень прозрачности" [14]. Правила регулирования, которые требуют, чтобы определенные системы ИИ объясняли сами себя, уже существуют и развивались на протяжении нескольких десятков лет. В частности, это системы кредитного рейтинга и системы по персональным данным GDPR [32].

В 1970 году в США был принят "Закон о справедливой кредитной отчетности", а затем в 1974 году "Закон о равных кредитных возможностях" [32]. Этот закон запретил "дискриминацию в кредитных решениях на основе расы, цвета кожи, религии, национального происхождения, пола, семейного положения, возраста (для взрослых), получение дохода от государственной помощи" [32]. Однако, в процессе предоставления объяснений по принятию решений по кредитам выяснилось, что "доступ к модели - это не то же самое, что понимание". Считается, что современные кредитные системы не основаны на сложных алгоритмах именно из-за изначального существования законодательного требования понятного объяснения [32]. "Существует явный компромисс между производительностью модели машинного обучения и его способность давать объяснимые и интерпретируемые прогнозы" [21].

Технологически "право на объяснение" для систем ИИ реализуется через направление Explainable AI (XAI) - объяснимый или прозрачный ИИ, который люди могут легко понять. "Объяснимый искусственный интеллект (XAI) — область, занимающаяся разработкой новых методов, объясняющих и интерпретирующих модели машинного обучения, получила огромное распространение в последние годы" [21].

Если законодательством или корпоративными документами компании будет установлено требование объяснимости принимаемых решений, то соответствующая система ИИ будет построена на более простых, понятных алгоритмах. Соответственно, компании, которые не будут связаны такими ограничениями, смогут использовать более сложные и эффективные алгоритмы, добиваясь конкурентного преимущества.

Уже известны примеры, когда компаниям удается не раскрывать алгоритмы если они официально признаны коммерческой тайной. Примером могут служить алгоритмы программы COMPAS, которая используется в США для определения профиля нарушителя [20]. Программа COMPAS оказалась предметом общественных расследований, были выдвинуты обвинения в предвзятости по признаку расы, но независимым экспертам не удалось получить доступ к модели. COMPAS является одним из примеров, насколько трудно соблюдать требование недискриминации при использовании больших данных и машинного обучения. В опроснике программы COMPAS даже нет вопроса о расе нарушителя, выдвигается гипотеза, что в данных, на которых обучалась программа, могла содержаться предвзятость и она сохранилась в общей модели [20].

Еще одним широко используемым методом является независимое квалифицированное тестирование, например Тест поставщика систем распознавания лиц (FRVT), проводимые Национальным институтом стандартов и технологий США. Алгоритмы представляются для тестирования в скомпилированном виде и тестируются как "black box" с интерфейсом тестирования C++. Разработчики теста считают, что "основным результатом оценки является то, что за последние пять лет (2013–2018 гг.) был достигнут значительный прирост точности" [16]. Поскольку не все компании готовы или фактически могут раскрывать и объяснять, как работают алгоритмы созданной ими ИИ системы, опция независимого тестирования, аудита, и сертифицирования является механизмом гарантии соблюдения принципов недискриминации, прозрачности, и подотчетности ИИ системы.

**Технологическая сингулярность и ее последствия**

Анализируя историю развития автономных систем управления корпорациями необходимо принимать во внимание понятие технологической сингулярности, которое применяется при анализе развития машинного искусственного интеллекта: "...взрыв все более высокого уровня интеллекта, поскольку каждое поколение машин в свою очередь создает более интеллектуальные машины" [9]. Как отмечает популяризатор технологической сингулярности Вернор Виндж (Vernor Vinge): "...если технологическая Сингулярность может случиться, она произойдет. ...когда прогресс в области автоматизации настолько убедителен, что принятие законов, или наличие обычаев, запрещающих такие вещи, просто гарантирует, что кто-то иначе получит их в первую очередь" [37]. Наиболее проработанным документом в области регулирования систем ИИ для гражданского и коммерческого использования является проект Акта об Искусственном интеллекте для стран Европейского Союза (далее Акт), поскольку некоторые другие страны (например, США и Китай) сосредоточились на законодательстве об автономных системах ИИ военного назначения. Европейская комиссия (ЕС) в проекте Акта уже предлагает установить "некоторые ограничения на свободу ведения бизнеса (статья 16) и свободу искусства и науки (статья 13) для обеспечения соблюдения приоритетных общественных интересов" [13]. В то же время, официальная стратегия нефтяной компании ADNOC (ОАЭ) до 2030 года уже предусматривает, что цифровой командный центр Panorama "в конечном итоге будет связан с клиентами и инвесторами, обеспечивая непрерывную интеграцию между заинтересованными сторонами" [34]. В 2021 году ADNOC получила награду в области отраслевых технологий за свой цифровой командный центр [1], что означает признание и дальнейшее распространение аналогичных технологий среди нефтегазовых компаний. Исходя из теории технологической сингулярности, компании, которые введут эффективный менеджмент на базе ИИ будут более эффективны и конкурентноспособны, соответственно у них будут ресурсы, чтобы внедрить еще более эффективную систему менеджмента. Даже если будут введены законодательные ограничения, компании могут пойти по уже существующему пути внутренней цифровой фабрики без официального назначения автономного директора.

Профессор Петрин предполагает, что "мы движемся к будущему, в котором «управление с помощью машины» должно будет в конечном итоге полностью заменить директоров и менеджеров-людей в качестве бизнес-лидеров" [27]. Наиболее очевидным случаем являются дочерние компании "выполняющие очень ограниченные функции" [2]. Такие компании "могут быть полностью автоматизированы" [2]. Будет ли это действительно одна объединенная система ИИ для управления компанией или несколько, зависит от того, чьи интересы такая система будет представлять. Вполне возможно, что разные группы акционеров выберут не одну и ту же систему ИИ, а разные. Петрин предполагает, что "крупные поставщики программного обеспечения будут предлагать услуги управленческого ИИ компаниям для продажи или аренды" [27].

Крупные компании - лидеры мирового рынка уже открыто объявляют, что собираются заменить значительную часть персонала, "который не работает с клиентами" системами ИИ [29]. В частности, компания IBM в ближайшие пять лет планирует заменить до 30% сотрудников поддерживающих функций системами ИИ [29]. Официально декларируемая позиция компаний и правительств состоит в том, что "ИИ должен быть инструментом для людей и силой добра в обществе с конечной целью повышения благосостояния людей" [13]. Тем не менее, как показывает пример IBM, может существовать значительный временной разрыв между изменением материального положения каждого конкретного сотрудника и увеличением общего общественного блага. Значительные сокращения квалифицированного персонала могут привести не только к социальным проблемам, но и к снижению научного и технического потенциала человеческой расы. Потеря дохода может быть компенсирована возросшими доходами от акций автономных компаний или программами минимального гарантированного дохода, но также необходимы программы для повышения квалификации или профессиональной переориентации сокращенных сотрудников. Эффективным решением может быть включение показателей поддержания качественных человеческих ресурсов в систему показателей ESG (экологическое, социальное и корпоративное управление). Таким образом, автономные компании будут стремиться не только к финансовым целям, но и к решению экологических, социальных, и управленческих проблем.

Математик Стивен Вольфрам (Stephen Wolfram) видит корень и решение проблемы конкуренции человека и ИИ в изменении системы образования: "мы медленно движемся к тому, чтобы люди получали образование в виде вычислительной парадигмы. И это хорошо, потому что, как я это вижу, вычисления станут центральным элементом почти в каждой области" [38]. Позиция Стивена Вольфрама представляется обоснованной, поскольку современные системы ИИ основаны именно на математике: "методы машинного обучения ...статистические методы, байесовские методы поиска и оптимизации." [13]. Если человечество продолжит двигаться по пути современного технического прогресса, основанного на вычислениях и математике, то ему придется перейти от культуры потребления к "культуре математики", что позволит создавать, управлять, и коммуницировать с системами ИИ.

При такой значительной передаче полномочий автономным системам неизбежно возникает вопрос ответственности за принимаемые такими системами решения. Профессор Петрин предполагает, что наиболее простой опцией для распределения прав и обязанностей будет уже применяемая система "ответственности производителя" [27]. Данный подход уже реализуется на практике. В проекте Акта об Искусственном интеллекте Еврокомиссия декларирует, что ответственность за работу систем ИИ может возлагаться на поставщика, импортёра, дистрибьютера, оператора или пользователя" [13]. На данный момент Еврокомиссия не рассматривает ИИ как юридически правоспособное лицо. В терминологии Акта системы ИИ рассматриваются как технический инструмент, продукт, поступающий на рынок, и ответственность за его применение возлагается всех участников цепочки создания стоимости такого продукта и его пользователей.

Еврокомиссия признает, что "автономное поведение ...может негативно повлиять на ряд основных прав, закрепленных в Хартии основных прав ЕС" [13]. Однако, автономные системы для управления компаниями в проекте Акта об Искусственном интеллекте даже не описаны. Системы же, которые обладают некоторыми начальными функциями по управлению компаниями - отнесены к системам ИИ высокого риска: если "ИИ предназначен для принятия решений о продвижении и прекращении трудовых договорных отношений, для распределения задач, а также для мониторинга и оценки эффективности и поведения лиц в таких отношениях" [13]. На данный момент для систем с высоким риском Еврокомиссия планирует установить строгие "требования высокого качества данных, документации и прослеживаемости, прозрачности, человеческого контроля, точности и надежности" [13]. В то время, как небольшие управленческие автономные системы ИИ уже разработаны и применяются [17], [28], [27], [6], Еврокомиссия только планирует создать для автономных систем "модель новой законодательной базы, реализуемой посредством внутренних контрольных проверок поставщиками" [13]. "После того, как поставщик проведет соответствующую оценку соответствия, он должен зарегистрировать эти автономные системы искусственного интеллекта с высоким риском в базе данных ЕС, которой будет управлять Комиссия, чтобы повысить общественную прозрачность и надзор, а также усилить постфактум надзор со стороны компетентных органов" [13]. Также, для систем высокого риска предусмотрено требование обязательного контроля человеком [13]. Если другие страны, подобно ЕС, не установят жесткое регулирование для автономных систем гражданского и коммерческого использования, то экономики таких стран и регионов могут получить ощутимое экономическое преимущество, которое в силу теории технологической сингулярности будет только увеличиваться.

**Заключение**

По мнению многих исследователей "ИИ должен стать центральным двигателем четвертой промышленной революции, которая навсегда изменит то, как мы живем на планете Земля. Это окажет значительное влияние на отдельных лиц и организации всех видов" [24]. Как справедливо отмечает профессор Армор "путь технологического развития — это траектория", однако проблема в том, что наиболее технически развитые компании на этой траектории уже значительно обогнали возможных конкурентов. Качественные автономные системы ИИ по аналогии с человеческим капиталом уже могут создавать добавленную стоимость в процессе производства [17]. На определенном этапе автономные системы могут оказаться эффективным инструментом для стран, регионов, и компаний с дефицитом человеческого капитала, уравнивая или предоставляя дополнительные шансы таким странам и компаниям в конкуренции на мировом рынке.